\documentclass[10pt,
amsmath,amssymb,aps,etoolbox,floatfix,
letterpaper,longbibliography,nofootinbib,
noshowpacs, preprintnumbers,reprint,showkeys,twocolumn]
{revtex4}

\usepackage{bm}       
\usepackage{caption}
\usepackage{color}
\usepackage{dcolumn}  
\usepackage{epsfig}
\usepackage{float}
\usepackage{gensymb}
\usepackage{graphicx} 
\usepackage{hyperref}

\newcommand{\beq}{\begin{equation}}

\newcommand{\eeq}{\end{equation}}

\begin{document}

\title{Frequency variation for {\emph {in vacuo}} photon propagation in the Standard-Model Extension}

\author{Jos\'e A. Helay\"el-Neto\textsuperscript{a}, 
Alessandro D.A.M. Spallicci\textsuperscript{b,c,d}\footnote{Corresponding author: spallicci@cnrs-orleans.fr\\ URL: http://wwwperso.lpc2e.cnrs.fr/$\sim$spallicci/}
}

\affiliation{
\mbox{\textsuperscript{a}Centro Brasileiro de Pesquisas F\'{\i}sicas (CBPF)}\\
\mbox{Departamento de Astrof\'{\i}sica, Cosmologia e Intera\c{c}\~{o}es Fundamentais (COSMO)}\\
\mbox {Rua Xavier Sigaud 150, 22290-180 Urca, Rio de Janeiro, RJ, Brasil}
\vskip1pt
\mbox{\textsuperscript{b}Universit\'e d'Orl\'eans}\\
\mbox{Observatoire des Sciences de l'Univers en r\'egion Centre (OSUC) UMS 3116} \\
\mbox{1A rue de la F\'{e}rollerie, 45071 Orl\'{e}ans, France}
\vskip1pt
\mbox{\textsuperscript{c}Universit\'e d'Orl\'eans}\\
\mbox{Collegium Sciences et Techniques (CoST), P\^ole de Physique}\\
\mbox{Rue de Chartres, 45100  Orl\'{e}ans, France} 
\vskip1pt
\mbox{\textsuperscript{d}Centre National de la Recherche Scientifique (CNRS)}\\
\mbox{Laboratoire de Physique et Chimie de l'Environnement et de l'Espace (LPC2E) UMR 7328}\\
\mbox {Campus CNRS, 3A Avenue de la Recherche Scientifique, 45071 Orl\'eans, France}
}

\date{3 July 2019}

\begin{abstract}
In the presence of Lorentz Symmetry Violation (LSV) associated with the Standard-Model Extension (SME), we have recently shown the non-conservation of the energy-momentum tensor of a light-wave crossing an Electro-Magnetic (EM) background field even when the latter and the LSV are constant. Incidentally, for a space-time dependent LSV, the presence of an EM field is not necessary. Herein, we infer that in a particle description, the energy non-conservation for a photon implies violation of frequency invariance {\it in vacuo}, giving rise to a red or blue shift. We discuss the potential consequences on cosmology.  
\end{abstract}


\maketitle

\section{Introduction} 

The Standard-Model (SM) describes through a Lagrangian three interactions among fundamental particles: Electro-Magnetic (EM), weak and strong. The SM is a very successful model but it neither includes massive neutrinos, nor incorporates the particles corresponding to a, yet to be found, dark universe. Furthermore, we remark that the photon is the only free massless particle in the SM.  

An attempt to extend the SM is Super-Symmetry (SuSy); see \cite{martin2016} for a review. This theory predicts the existence of new particles that are not included in the SM. Anyway, the physics we describe herein is valid also in absence of a SuSy scenario. In this respect, the role of SuSy is solely the provision of a microscopic origin of the LSV.  

The SM is assumed to be Lorentz Symmetry (LoSy)\footnote{Poincar\'e has equally contributed to the establishment of these symmetries.}  invariant. This prediction is likely valid only up to certain energy scales beyond which a LoSy Violation (LSV) might occur. There is a general framework known as the SM Extension (SME) \cite{colladaykostelecky1997,colladaykostelecky1998,liberati2013}, that allows us to test the low-energy manifestations of LSV. 

In two recent works \cite{bodshnsp2017,bodshnsp2018} on the SME, we have considered violations of LoSy, differing in the handedness of the Charge conjugation-Parity-Time reversal  
(CPT) symmetry and in whether considering the impact of photinos on photon propagation.  We came up with four classes. For the CPT-odd classes ($k^{\rm AF}_{\alpha}$ breaking vector) associated with the Carroll-Field-Jackiw (CFJ) model, the dispersion relations (DRs) and the Lagrangian show for the photon an effective mass, gauge-invariant, and proportional to 
$|{\vec k}^{\rm AF}|$. The group velocity exhibits a deviation from the speed of light $c$. The deviation depends on the inverse of the frequency squared, as predicted by de Broglie \cite{db40}.  
For the CPT-even classes ($k_{\rm F}^{\alpha\nu\rho\sigma}$ breaking tensor), when the photino is considered, the DRs display also a massive behaviour inversely proportional to a coefficient in the Lagrangian and to a term linearly dependent on 
$k_{\rm F}^{\alpha\nu\rho\sigma}$. All DRs feature an angular dependence and lack LoSy invariance.
Complex or simply imaginary frequencies and super-luminal speeds may appear in defined cases. Furthermore, we have shown the emergence of birefringence.
Finally, for both CPT sectors, we have pointed out the non-conservation of the photon energy-momentum tensor {\it in vacuo} \cite{bodshnsp2018}. 

Hereafter, we deal with the latter result and give an order of magnitude of the energy change that light would undergo through propagation in a LSV universe. The energy variations, if losses, would translate into frequency damping if the excitation were a photon.  
Generally, the wave-particle correspondence, even for a single photon \cite{aspect-grangier-1987}, leads us to consider that the non-conservation of energy corresponds to a photon energy variation and thereby a red or a blue shift. 

Before stepping into the equations, we intend to present the physical reason for why the non-conservation arises even in case of a constant EM background and of a constant LSV breaking vector (the breaking tensor appears either under a derivative or coupled to a derivative of the EM background). 

We recall that the CFJ equations of motion and the action are gauge-invariant but they originate from a Lagrangian density which is not gauge-invariant. Indeed, the gauge dependence of the Lagrangian density is a surface term to be neglected in the action. 
Conversely, gauge invariance is not acquired when processing the Lagrangian density of the classical massive electromagnetism of de Broglie-Proca.

Concerning the non-conservation, the action contains a contribution 
$\epsilon^{\kappa\lambda\mu\nu}k^{\rm AF}_\kappa A_\lambda F_{\mu\nu}$, such that even if the 
EM background is constant, the corresponding background four-potential is not, $A_\beta = x^\alpha F_{\alpha\beta}$. Thereby, there is an explicit $x^\alpha$ dependence at the level of the Lagrangian. This determines a source of energy-momentum non-conservation, according to the Noether theorem. Otherwise put, there is an exchange of energy-momentum between the photon and the EM background. 
The latter is external to the system and does not follow the dynamics dictated by the photon action. 

We also remark that the four-curl of $k^{\rm AF}$ is zero. This guarantees gauge invariance of the action and, in a simply connected space, $k^{\rm AF}$  may be expressed as the four-gradient of a scalar function. 

\section{Energy-momentum non-conservation} 

Our most general scenario is composed of $k^{\rm AF}_\alpha$ and $k_{\rm F}^{\alpha\nu\rho\sigma}$; $f_{\alpha\nu}$ represents the photon field and $a_\nu$ is the four-potential; $F_{\alpha\nu}$ the EM background field, $j^{\nu}$ the external current independent of the latter. The symbol * stands for the dual field. Starting from the field equation \cite{bodshnsp2018} in SI units ($\mu_0 = 4\pi \times 10^{-7}$ NA$^{-2}$), where we used  
$\partial_\alpha k^{\rm AF}_{\nu} - \partial_\nu k^{\rm AF}_{\alpha} = 0$ for the virtue of gauge invariance, and where 
${\cal F}^{\mu\nu} = F^{\mu\nu} + f^{\mu\nu}$ is the total field

\beq
\partial_{\alpha}{\cal F}^{\alpha\nu}+ k^{\rm AF}_{\alpha}\ ^{*}{\cal F}^{\alpha\nu}+\left(\partial_{\alpha}k_{\rm F}^{\alpha\nu\kappa\lambda}\right){\cal F}_{\kappa\lambda}
+k_{\rm F}^{\alpha\nu\kappa\lambda}\partial_{\alpha}F_{\kappa\lambda} 
= \mu_0 j^{\nu}~,
\eeq
and adopting the identities indicated in \cite{bodshnsp2018},
we worked out the photon energy-momentum tensor

\begin{align}
\theta_{\ \rho}^{\alpha} & = \frac{1}{\mu_0} \left( f^{\alpha\nu}f_{\nu\rho}
+ \frac{1}{4} \delta_{\rho}^{\alpha}f^{2}
- \frac{1}{2} k^{\rm AF}_{\rho}\ ^{*}f^{\alpha\nu}a_{\nu}+\right.\nonumber \\
 & \ \ \left. k_{\rm F}^{\alpha\nu\kappa\lambda}f_{\kappa\lambda}f_{\nu\rho}
+\frac{1}{4} \delta_{\rho}^{\alpha}k_{\rm F}^{\kappa\lambda\alpha\beta}f_{\kappa\lambda}f_{\alpha\beta}\right )~,
\label{pemt}
\end{align}
and its non-conservation
\begin{align}
\partial_{\alpha}\theta_{\ \rho}^{\alpha} & = j^{\nu}f_{\nu\rho}
- \frac{1}{\mu_0} \left [\left(\partial_{\alpha}F^{\alpha\nu}\right)f_{\nu\rho}
+ k^{\rm AF}_{\alpha}\ ^{*}F^{\alpha\nu}f_{\nu\rho} + \right.\nonumber \\
 & \frac{1}{2}\left(\partial_{\alpha}k^{\rm AF}_{\rho}\right)\ ^{*}f^{\alpha\nu}a_{\nu}
-  \frac{1}{4}\left(\partial_{\rho}k_{\rm F}^{\alpha\nu\kappa\lambda}\right)f_{\alpha\nu}f_{\kappa\lambda} + \nonumber \\
 & \left. \left(\partial_{\alpha}k_{\rm F}^{\alpha\nu\kappa\lambda}\right)F_{\kappa\lambda}f_{\nu\rho}
+ k_{\rm F}^{\alpha\nu\kappa\lambda}\left(\partial_{\alpha}F_{\kappa\lambda}\right)f_{\nu\rho}\right]~.
\label{pemt-nc}
\end{align}

As mentioned, although derived in a SuSy framework embedding LSV, Eqs. (\ref{pemt},\ref{pemt-nc}) are applicable without any reference to SuSy. Few remarks appear necessary for appreciating Eqs. (\ref{pemt},\ref{pemt-nc}). 

\begin{itemize}  
\item{The right-hand side of Eq. (\ref{pemt-nc}) exhibits all types of terms that describe the exchange of energy between the photon,
the LSV parameters, the EM background field and the external current, taking into account an $x^\alpha$-dependence of the LSV parameters and of the EM background field. }
\item{In Eq. (\ref{pemt-nc}), the first two right-hand side terms are purely Maxwellian.}
\item {The energy-momentum tensor in Eq. (\ref{pemt}) loses its symmetry, and thereby $\theta_{0i}\neq \theta_{i0}$. This tells us that 
the momentum density $\theta_{0i}$ does not correspond any longer to the extended Poynting vector $\theta_{i0}$. Setting $\rho= i$ in Eq. (\ref{pemt-nc}), we have $\partial_{\alpha}\theta_{\ \rho}^{\alpha} = \partial_{0}\theta_{\ i}^{0} + 
\partial_{j}\theta_{\ i}^{j} = j^\nu f_{\nu i} + .... = j^0 f_{0 i} + j^k f_{k i} + ....$, so that the density of the Lorentz force 
appears at the right-hand side. Therefore, we {interpret} $\theta_{\ i}^{0}$ as the momentum density of the wave (the time derivative of the momentum provides the force).  
}
\item{We return to a comment made in the {\it Introduction}. In Eq. (\ref{pemt-nc}) the term $k^{\rm AF}_{\alpha}\ ^{*}F^{\alpha\nu}f_{\nu\rho}$ is space-time independent. Indeed, $k^{\rm AF}_{\alpha}$ from the CFJ Lagrangian \cite{chernsimons1974} depends on the four-potential. By splitting the total field in background and photon fields, an explicit dependence on the EM background potentials appears now in the CFJ Lagrangian \cite{cafija90}. But, if the background field is constant, the background potential must necessarily show a linear dependence on $x_\mu$ and translation invariance of the Lagrangian is thereby lost.}
\item{The term $k_{\rm F}^{\alpha\nu\kappa\lambda}\left(\partial_{\alpha}F_{\kappa\lambda}\right)f_{\nu\rho}$, even if 
$k_{\rm F}^{\alpha\nu\kappa\lambda}$ is constant, breaks translation invariance due to the space-time dependence of the EM background field.} 
\item{We finally notice that there is energy non-conservation even in absence of an EM background field and of an external current. This is due to the presence of LSV space-time dependent terms.}
\end{itemize}

Since we are focusing on energy, we can tailor Eq. (\ref{pemt-nc}) to our needs, and thereby we set $\rho = 0$. Due to the absence of diagonal terms in the EM field tensor, where this is applicable, $\nu$ takes only spatial values $i$. We have 

\begin{align}
\partial_{\alpha}\theta_{\ 0}^{\alpha} & = j^{i}f_{i 0}
- \frac{1}{\mu_0} \left [\left(\partial_{\alpha}F^{\alpha i}\right)f_{i 0}
+ k^{\rm AF}_{\alpha}\ ^{*}F^{\alpha i}f_{i 0} + \right.\nonumber \\
 & \frac{1}{2}\left(\partial_{\alpha}k^{\rm AF}_{0}\right)\ ^{*}f^{\alpha\nu}a_{\nu}
-  \frac{1}{4}\left(\partial_{0}k_{\rm F}^{\alpha i\kappa\lambda}\right)f_{\alpha\nu}f_{\kappa\lambda} + \nonumber \\
 & \left. \left(\partial_{\alpha}k_{\rm F}^{\alpha i\kappa\lambda}\right)F_{\kappa\lambda}f_{i 0}
+ k_{\rm F}^{\alpha i \kappa\lambda}\left(\partial_{\alpha}F_{\kappa\lambda}\right)f_{i 0}\right]~.
\label{pemt-nc-0}
\end{align}

{\renewcommand{\arraystretch}{1.2}
\begin{table}[h]
\centering
\caption
{\small
Upper limits of the LSV parameters (the last value is in SI units): 
$^{\rm a}$Energy shifts in the spectrum of the hydrogen atom \cite{gomesmalta2016};  
$^{\rm b}$Rotation of the polarisation of light in resonant cavities \cite{gomesmalta2016}; 
$^{\rm c,e}$Astrophysical observations \cite{kosteleckyrussell2011}. Such estimates are close to the Heisenberg limit on the smallest measurable energy or mass or length for a given time $t$, set equal to the age of the universe; 
$^{\rm d}$Rotation in the polarisation of light in resonant cavities \cite{gomesmalta2016}.  
$^{\rm f}$Typical value \cite{kosteleckyrussell2011}.  
}
\begin{tabular}{|c|c|}
\hline
$|{\vec k}^{\rm AF}|$~~~~~$^{\rm a}$                         & 
$< 10^{-10}$ eV $ = 1.6 \times 10^{-29}$ J$;~ 5.1 \times 10^{-4}$ m$^{-1}$         \\ \hline        
$|{\vec k }^{\rm AF}|$~~~~~$^{\rm b}$                         & 
$< 8\times 10^{-14}$ eV $ = 1.3 \times 10^{-32}$ J$;~ 4.1 \times 10^{-7}$ m$^{-1}$ \\ \hline 
$|{\vec k}^{\rm AF}|$~~~~~$^{\rm c}$                         & 
$< 10^{-34}$ eV $ = 1.6 \times 10^{-53}$ J$;~ 5.1 \times 10^{-28}$ m$^{-1}$        \\ \hline   
$k^{\rm AF}_0$~~~~~~$^{\rm d}$                               &            
$< 10^{-16}$ eV $ = 1.6 \times 10^{-35}$ J$;~ 5.1 \times 10^{-10}$ m$^{-1}$         \\ \hline  
$k^{\rm AF}_0$~~~~~~$^{\rm e}$                               &            
$< 10^{-34}$ eV $ = 1.6 \times 10^{-53}$ J$;~ 5.1 \times 10^{-28}$ m$^{-1}$         \\ \hline  
$k_{\rm F}$~~~~~~$^{\rm f}$                                 &
 $\simeq 10^{-17}$                                 \\ \hline    
\end{tabular}
\label{table}
\end{table}
}

Table \ref{table} provides the upper limits of the LSV terms.

\section{Sizing the EM background field}
For the magnetic fields, we refer to \cite{alves-ferriere-2018,ferriere-2018}. 

\paragraph{Spatial dependence of the magnetic field in the Milky Way.}

The inter-stellar magnetic field in the Milky Way has a strength of around 500 pT.
It has regular and fluctuating components of comparable strengths. 
The Galactic disk contains the regular field, which is approximately horizontal and parallel, being spirally shaped with a generally small opening angle of about $p = 10^{\circ}$.
In cylindrical coordinates,
$B \simeq B_r \cdot e_r + B_\phi \cdot e_\phi$, with $B_r = B_\phi \tan p$. 

In the Galactic halo, the regular field is not horizontal, probably holding an X-shape, as observed in spiral galaxies.
The fluctuating field varies over a whole range of spatial scales, from 100 parsecs down to very small scales.

\paragraph{Time dependence of the magnetic field in the Milky Way.} 

The regular field evolves over very long time scales such as 1 Gyr.
It likely increases exponentially in time until an equi-partition with kinetic energy is achieved. At that point, it saturates.
Indeed the time derivative of the magnetic field obeys an equation containing spatial derivatives of $B$ which coefficients of are independent of $B$ until the counter-reaction of the inter-stellar fluid small-scale turbulent motion comes into play. 
Physically, the galaxy large-scale shearing of the poloidal field generates an azimuthal field, which in turn generates a poloidal field. 
The solution of this type of equation is indeed exponential in time. It is a dynamo mechanism.

The fluctuating field varies over much shorter time scales, probably 1 Myr.

\paragraph{Other galaxies.}  

External galaxies also possess inter-stellar magnetic fields with strengths of several hundred pT.
While, in spiral galaxies the fields resemble those in our own Galaxy, there is absence of the regular component in elliptical galaxies, and solely fluctuating components are present.

\paragraph{Inter-galactic space.}  

No certain conclusion can be drawn on the Inter-Galactic Medium (IGM). 
The medium between galaxies inside a cluster of galaxies hosts a fluctuating field with a typical strength of a few nT.
The IGM outside of clusters of galaxies may also contain magnetic fields. Claims have been laid to the detection of such fields, but a confirmation is missing.

\paragraph{Electric field.} 

The inter-stellar and inter-galactic media are good electric conductors, such that magnetic fields are frozen in the plasma. Thereby, the electric field is given by
${\vec E} \propto {\vec v}_p \times {\vec B}$,
where $v_p$ is the plasma velocity.
In general, $v_p \ll c$, thus $E \ll B$ and thereby neglected herein. This assumption may not hold locally, and photons may pass through intense electric fields.

\section{Sizing the energy non-conservation}

In Eq. (\ref{pemt-nc-0}), we neglect the tensorial perturbation, $k_{\rm F}$ on the basis that is less likely to condensate, taking an expectation value different from zero, in contrast to the vectorial CFJ perturbation. If we consider that SuSy is a viable path beyond the SM, in \cite{bebegahn2013,bebegahnle2015}, it is shown that $k_{\rm F}$ emerges as the product of multiple SuSy condensates in contrast to 
$k_{\rm AF}$, which consists of a single SuSy condensate; therefore the latter is dominating as compared to $k_{\rm F}$. 

On the other hand, independently from the considerations based on SuSy, we justify 
neglecting the $k_{\rm F}$ term on other grounds. This term is quadratic in the field strength and in the frequency. The CFJ term instead contains a single derivative and thereby it is
linear in the frequency. If we confine our investigation to low frequencies, as we do here, it is
reasonable that the $k_{\rm AF}$ term yields the dominating contribution. Instead, for very high frequencies,
we expect that the $k_{F}$ term dominates.

Further, we suppose that $k^{\rm AF}_0$ is constant. We thus get
 
\beq
\partial_{\alpha}\theta_{\ 0}^{\alpha} = j^{i}f_{i 0} -\frac{1}{\mu_0} \left [\left(\partial_{\alpha}F^{\alpha i}\right)f_{i 0}
+ k^{\rm AF}_{\alpha}\ ^{*}F^{\alpha i}f_{i 0} \right ]~.
\label{pemt-nc-0-new}
\eeq

We are interested in the change of energy along the line of sight $x$ where the photon path lies.
We intend to render the terms in Eq. (\ref{pemt-nc-0-new}) explicit. In absence of an electric field, only the spatial components of the EM background field tensor are present as well as the mixed space-time components of the dual EM background field tensor, that is the magnetic field. We suppose also the absence of an external current. Equation (\ref{pemt-nc-0-new}) is approximated by ($\vec e$ and $\vec b$ are the electric and magnetic field of the photon, respectively)  

\begin{align}
& \frac{1}{2}\frac{\partial}{\partial t}\left(\epsilon_0 {\vec e}^{\!~2} - \frac{k^{\rm AF}_{0}}{\mu_0c}{\vec e}\cdot{\vec a} + \frac{{\vec b}^{\!~2}}{\mu_0}\right) 
+ \frac{1}{\mu_0} \frac{\partial}{\partial x} \left({\vec e} \times {\vec b}\right)_x = \nonumber \\
& -\frac{c}{\mu_0} \left [\left(\partial_{x}F^{xi}\right)f_{i 0}
+ k^{\rm AF}_{0}\ ^{*}F^{0 i}f_{i 0} \right ] = \nonumber \\ 
& -\frac{c}{\mu_0} \left [\underbrace{\partial_{x}F^{xi}}_{\rm First~term}
+ \underbrace{k^{\rm AF}_{0}\ ^{*}F^{0 i}}_{\rm Second~term}\right ]f_{i 0}~.
\label{pemt-nc-0-new-expl}
\end{align} 

The dimensions in Eq. (\ref{pemt-nc-0-new-expl}) are Jm$^{-3}$s$^{-1}$. 
The left-hand side of Eq. (\ref{pemt-nc-0-new-expl}) corresponds to the expansion of $\partial_{\alpha}\theta_{\ 0}^{\alpha}$, with $\theta_{\ 0}^{\alpha}$ given by Eq. (\ref{pemt}) where the $k_{\rm F}$ contribution has been neglected for the reasons previously stated.

We {assume} the absence of IGM magnetic field fluctuations over long time scales, that amounts to considering only the time fluctuations in the emitting galaxy and in our Milky {Way}, estimated as $10^{21}$ m in size. The first term is estimated as $5\times 10^{-10}/10^{21}$ Tm$^{-1}$, and thereby dropped henceforth. Under all these assumptions, the energy variation comes chiefly from the second term. 

The $k^{\rm AF}_{0}$ component of the LSV vector extends to the entire universe and thus it is not confined to a limited region. We need to integrate over the light travel time. For a source at 
$z = 0.5$, the look-back time is $t_{LB}= {5} \times 10^9$ yr = { $1.57 \times 10^{17}$} s {\cite{wright-2006}}, having taken a somewhat mean value among different values of {the} cosmological parameters ($H_0 = 70$ km/s per Mpc Hubble-Humason constant, $\Omega_m {= 0.3}$ matter density, $\Omega_\Lambda {= 0.7}$ dark energy density). 
We set an arbitrary safe margin $\varrho$, defined as positive, to take into account that the many magnetic fields, estimated at $B = 5 \times 10^{-10} - 5 \times 10^{-9}$ T each, and crossed by light from the source to us, have likely different orientations and partly compensate their effects on the wave energy\footnote{We have not considered the potential presence of a strong magnetic field at the source.}. 

Thus, the energy density {change of the wave} due to the second term, $\Delta${ \textsf E}, is ($B = 2.75$ nT)       

\beq
|\Delta {\textsf E}|_{z=0.5} = \frac{c}{\mu_0} k^{\rm AF}_0 |B f_{i 0}| {\varrho}~t_{LB} \approx  
{  1.02 \times 10^{23}}~k^{\rm AF}_0 {\varrho} |f_{i 0}|~.
\eeq

For $h=6.626 \times 10^{-34}$ Js, the frequency change $\Delta\nu$ is 

\beq
|\Delta\nu|_{z=0.5} = \frac{  1.02 \times 10^{23}}{h}k^{\rm AF}_0 {\varrho} {|f_{i 0}|} = { 1.55 \times 10^{56}}~k^{\rm AF}_0 {\varrho} |f_{i 0}|~.
\label{estimate1}
\eeq

We now need to compute $|f_{i 0}| = |{ {\vec e}}|/c$, the electric field of the photons. We consider the Maxwellian - in first approximation - classic intensity $I = \epsilon_0 c {e^2} = \epsilon_0 c^3 f_{i0}^2$ ($c{ b} = {e}$).

The frequency 
$\nu = 4.86 \times 10^{14}$ Hz corresponds to the Silicon absorption line at 6150 \AA, of 1a Super-Nova (SN) type. 
The monochromatic AB magnitude is based on flux measurements that are calibrated in absolute units \cite{oke-gunn-1983}. It is defined as the logarithm of a spectral flux density\footnote{The spectral flux density is the quantity that describes the rate at which energy density is transferred by EM radiation per unit wavelength.} $SFD$ with the usual scaling of astronomical magnitudes and about 3631 Jy as zero-point\footnote{In radio-astronomy, the jansky is a non-SI unit of spectral flux density equivalent to $10^{-26}$ W m$^{-2}$ Hz$^{-1}$.} 

\beq
m_{\rm AB} = -2.5 \log_{10}SFD - 48.6~,
\eeq
in cgs units. For $m_{\rm AB} = - 19$ {(appropriate for SN Ia around the maximal light in this wave band)}, we get $SFD = {1.44 \times} 10^{-15}$ Js$^{-1}$~Hz$^{-1}$~m$^{-2}$ having been converted to SI units. We integrate over the frequency width of a bin, that is 30 \AA~or\footnote{For the frequency width, we have computed $\frac{\Delta\lambda}{\lambda}=\frac{\Delta \nu}{\nu}$. } 2.37 THz and get $I= {3.4 } \times 10^{-3}$ Js$^{-1}$~m$^{-2}$. 
For $\epsilon_0 = 8.85 \times 10^{-12}$ Fm$^{-1}$, we have  

\beq
f_{i 0} = \sqrt{\frac{I}{\epsilon_0c^3}} = {3.79} \times 10^{-9} {\rm Vsm}^{-2}~. 
\label{estimate2}
\eeq

Finally, from Eq. (\ref{estimate1}), we get
 
\beq
|\Delta\nu|_{z=0.5}^{\nu = 486 {\rm THz}} = { 5.87 \times 10^{47}}~k^{\rm AF}_0 {\varrho} ~.
\label{estimate3new}
\eeq

The large range of values of $k^{\rm AF}_0$ and $\varrho$ render the range of values for the estimate in Eq. (\ref{estimate3new}) also large. 
We recall that $z = \Delta \nu/\nu_o$ where $\Delta \nu = \nu_e - \nu_o$ is the difference between the observed $\nu_o$ and emitted 
$\nu_e$ frequencies. For $z_c = 0.5$, where $z_c$ is the redshift due to the expansion of the universe, 
$|\Delta\nu|_{z=0.5}^{\nu = 486 {\rm THz}} = 1.62$ THz. We ask whether we can get a similar value for $z_{LSV}$, the shift due to LSV.

We consider two numerical applications.
For $k^{\rm AF}_0 = 10^{-10}$m$^{-1}$, Tab. \ref{table}, and ${ \varrho} { ~\approx 10^{-23}}${, which represents an extreme misalignment of the magnetic fields}, or 
for $k^{\rm AF}_0 = {5.1 \times 10^{-28}}$m$^{-1}$, Tab. \ref{table}, and ${ \varrho} { ~ \approx 10^{-6}}$, we get 
$|\Delta\nu|_{z=0.5}^{\nu = 486 {\rm THz}}$ 
in the range of $10^{14}$ Hz. This would strongly influence the measurement of the redshift due to the expansion of the universe, since $z_{LSV}$ would be comparable to $z_c$. Instead, combining the astrophysical upper limit on the size of $k^{\rm AF}_0$ with a value of ${ \varrho} \ll 10^{-7}$, it will conversely produce a small effect.

\vskip 10pt
  
\section{Impact on cosmology}
 
We have determined an expression for an LSV frequency shift. The size of the effect may be negligible for cosmology, but just of relevance for the foundations of physics. Nevertheless, the rough estimates in the previous section seem to point to a large impact. Here below, we consider that the LSV shift takes a value large enough to be considered, and thereby to be superposed to the cosmological redshift. Which interpretation should we adopt in analysing spectra from distant sources?  

The parameter $z$ is given by $z = \Delta/\lambda_e$ where $\Delta {\lambda} = \lambda_o - \lambda_e$ is the difference between the observed $\lambda_o$ and emitted $\lambda_e$ wavelengths. Expansion causes $\lambda_e$ to stretch to $\lambda_c$ that is $\lambda_c = (1+z_c)\lambda_e$. The wavelength $\lambda_c$ could be further stretched or shrunk for the supposed LSV shift to $\lambda_o = (1+z_{LSV})\lambda_c  = (1+z_{LSV}) (1+z_c) \lambda_e$. But since $\lambda_o = (1+z)\lambda_e$, finally we have 

\begin{align}
1 +  z = (1+z_{LSV}) (1+z_c) = 1 + z_{LSV} + z_c  + {\cal O}(z^2)~. 
\end{align}

A reverse estimate process would instead set an error of the redshift measurement and assess upper limits to the LSV parameters.  

\section{Conclusions, discussion and perspectives}
 
We have introduced a new frequency shift for {\it in vacuo} propagation of a photon in a LSV scenario. 
The physical situation is as follows. We have neglected time variations of the LSV breaking terms and of the magnetic fields. Thus, the time averaging of the LSV shift differs from zero. Along the line of sight, the space averaging is also never zero, unless – obviously – there isn't any LoSy breaking, or the magnetic field vectors perfectly cancel one another. But, for the observer, there is an angular dependence of the LSV frequency shift, due to the LSV itself. For each direction, there is a value of $k^{\rm AF}_0$ and of $\varrho$, and thereby a direction-dependent LSV shift. The issue is whether the LSV shift is large enough to have an impact on the observations. 
  
We certainly need to put stringent model independent observational and experimental upper limits to $z_{LSV}$ through constraints on the LSV parameters and on the EM field values. We question whether the sign of $z_{LSV}$, and thereby a red or blue shift, could not be determined {\it a priori} on the basis of perturbation theory. Undoubtedly, the orientations and scale lengths of the LSV parameters, as well as the photon path crossing multiple background EM fields differently oriented, render this shift very dependent on the trajectory. 

We remark that the 
discrepancy between the luminosity distance derived with standard cosmology and the data, nowadays mostly explained by assuming dark energy, should be reviewed in light of this additional frequency shift. 

Classic electromagnetism has been well tested, as general relativity. This has not impeded the proposition of alternative formulations of gravitation during last century, and lately to circumvent the need of dark matter and dark energy. We point out that revisiting astrophysical data with non-Maxwellian electromagnetism opens the door to radically new interpretations. 

For instance, if we suppose that a static source bursts, and that at start it emits higher frequencies than at the end, this may mimic a time dilation effect from a receding source, if massive photons are considered. Indeed, for the CPT-odd handedness classes associated with LSV which entail massive photons, the deviation from $c$ of the group velocity is inversely proportional to the square of the frequency. Thereby, the photons emitted towards the end of the burst will employ more time than the initial photons to reach an observer. Incidentally, the dependence of the group velocity on the frequency allows us to set upper limits on the photon mass from Fast Radio Bursts (FRBs) \cite{boelmasasgsp2016,wuetal2016b,boelmasasgsp2017,bebosp2017,shao-zang-2017}. 

Generally, there is a continuous interest for testing non-Maxwellian electromagnetism, let it be massive or non-linear. The official upper limit on the photon mass is 10$^{-54}$ kg \cite{tanabashietal2018}\footnote{de Broglie estimated the photon mass lower than 10$^{-53}$ kg already in 1923 with a strike of genius \cite{db23}.}, but see \cite{retinospalliccivaivads2016} for comments on the reliability of such a limit and for an experiment with solar wind satellite data. While opening a new low-frequency radio-astronomy window with a swarm of nano-satellites would be desirable \cite{bebosp2017}, terrestrial experiments are faster to implement \cite{rosato-2019}. Among the non-linear effects, the last one to be detected is photon-photon scattering at CERN \cite{aaboud-et-al-2017}. 

\section*{Acknowledgements} 
The authors thank K. Ferri{\`e}re (Toulouse) for the information on galactic and inter-galactic EM fields; M. L\'opez Corredoira (La Laguna) and M. Nicholl (Edinburgh) for comments on SNs. Thanks are also due to L.P.R. Ospedal (CBPF) for long and fruitful discussions on aspects of LSV, to M. Makler (CBPF) for remarks on dark energy, and to J. Sales de Lima (S\~{a}o Paulo) for his 
interest. The anonymous referee provided constructive remarks.

\bibliographystyle{unsrt}

\end{document}